\documentclass[12pt,twoside,a4paper]{article}
\usepackage{amsmath,amssymb,latexsym,theorem,natbib,epsfig,color,subfigure,footmisc,bbm}
\usepackage{multirow,graphicx,array,rotating,url,wrapfig}  
\usepackage{epstopdf,afterpage,wasysym}
\usepackage{verbatim}
\usepackage[normalem]{ulem}
\def\crps{\mathop{\hbox{\rm CRPS}}}
\def\crpss{\mathop{\hbox{\rm CRPSS}}}
\def\ri{\mathop{\hbox{\rm RI}}}

\numberwithin{equation}{section}

\title{Machine learning-based probabilistic forecasting of solar irradiance in Chile}

\author{S\'andor Baran$^{1,*}$, Julio C. Mar\'\i n$^{2,3}$, Omar Cuevas$^{3,4}$,  Mailiu D\'{\i}az$^5$, \\Marianna Szab\'o$^1$, Orietta Nicolis$^5$  and M\'aria Lakatos$^1$  \vspace*{0.5cm}\\
  {\small $^1$Faculty of Informatics, University of Debrecen, Hungary}\\
   {\small $^2$Department of Meteorology, University of Valpara\'\i so, Chile}\\
   {\small $^3$Center for Atmospheric Studies and Climate Change (CEACC), University of Valpara\'\i so, Chile}\\
  {\small $^4$Institute of Physics and Astronomy, University of Valpara\'\i so, Chile}\\
     {\small $^5$Faculty of Engineering,  Andr\'es Bello University, Chile}\\
}

\date{}

\begin{document}

\maketitle

\footnotetext[1]{Corresponding author: \url{baran.sandor@inf.unideb.hu}}

\begin{abstract}
By the end of 2023, renewable sources cover 63.4\,\% of the total electric power demand of Chile, and in line with the global trend, photovoltaic (PV) power shows the most dynamic increase. Although Chile's Atacama Desert is considered the sunniest place on Earth, PV power production, even in this area, can be highly volatile. Successful integration of PV energy into the country's power grid requires accurate short-term PV power forecasts, which can be obtained from predictions of solar irradiance and related weather quantities.
Nowadays, in weather forecasting, the state-of-the-art approach is the use of ensemble forecasts based on multiple runs of numerical weather prediction models. However, ensemble forecasts still tend to be uncalibrated or biased, thus requiring some form of post-processing.
The present work investigates probabilistic forecasts of solar irradiance for Regions III and IV in Chile. For this reason, 8-member short-term ensemble forecasts of solar irradiance for calendar year 2021 are generated using the Weather Research and Forecasting (WRF) model,  which are then calibrated using the benchmark ensemble model output statistics (EMOS) method based on a censored Gaussian law, and its machine learning-based distributional regression network (DRN) counterpart. Furthermore, we also propose a neural network-based post-processing method resulting in improved 8-member ensemble predictions.
All forecasts are evaluated against station observations for 30 locations, and the skill of post-processed predictions is compared to the raw WRF ensemble. Our case study confirms that all studied post-processing methods substantially improve both the calibration of probabilistic- and the accuracy of point forecasts. Among the methods tested, the corrected ensemble exhibits the best overall performance. Additionally, the DRN model generally outperforms the corresponding EMOS approach.

\end{abstract}

\section{Introduction}
\label{sec1}

According to the latest report of the International Renewable Energy Agency \citep{irena24}, the largest ever increase in renewable power capacity was observed in 2023, nearly 75\,\% of which was newly installed solar energy. As a result, by the end of 2023, the renewable energy share had reached 43\,\% of the global installed power capacity, and this ratio is even higher in South America (71.4\,\%). In particular, renewable sources cover 63.4\,\% of the total electric power demand of Chile, 39.7\,\% of which comes from photovoltaic (PV) energy. In line with the global trend, with the addition of 1949 megawatts, in 2023, PV power accounted for the most substantial increase of 30.4\,\%.

Although Chile’s Atacama Desert is considered the sunniest place on Earth, with the highest long-term solar irradiance \citep{rmf15}, PV power production can be highly volatile, which raises a strong demand for accurate PV power forecasts from power grid operators. A standard approach to PV power forecasting is to consider global horizontal irradiance (GHI) forecasts (and possibly forecasts of other weather variables) and convert them to PV power with the help of a model chain, see e.g. \citet{my22} or \citet{hkl24}. In the present study, we concentrate on solar irradiance predictions, as they are highly correlated with the PV model chain outputs.

Solar irradiance forecasts are obtained traditionally as outputs of numerical weather prediction models (NWP), which describe the behaviour of the atmosphere with the help of partial differential equations. The state-of-the-art approach is to run these models simultaneously with various initial conditions and/or parametrizations, resulting in a probabilistic prediction as an ensemble forecast \citep{btb15,b18a}. Nowadays, all major weather centers operate their ensemble prediction systems (EPSs); one of the most prominent is the Integrated Forecast System (IFS) of the European Centre for Medium-Range Weather Forecasts (ECMWF) providing 51-member medium-range ensemble forecasts at 9 km resolution and 101-member extended range forecasts at 32 km resolution \citep{ecmwf23}. Nevertheless, in the last few years, NWP models got strong competitors in machine learning-based, fully data-driven forecasts such as the Pangu-Weather \citep{bxz23} or the more recent ECMWF Artificial Intelligence/Integrated Forecasting System \citep[AIFS;][]{lac24}, which is the first to issue AI-based ensemble predictions.

Despite the efforts devoted to improving the EPSs, ensemble forecasts might exhibit deficiencies such as bias or lack of calibration, thus requiring some form of statistical post-processing \citep{b18b}. In the last decades, many post-processing methods have been suggested for a wide spectrum of weather quantities; for an overview see e.g. \citet{vbd21} or \citet{sl22}. Among these parametric approaches are the ensemble model output statistics  \citep[EMOS;][]{grwg05} or distributional regression networks \citep[DRN;][]{rl18}, which provide full predictive distributions in the form of a single parametric law. In the EMOS method, the parameters of the predictive distribution depend on the forecast ensemble via appropriate link functions, whereas in the DRN approach, one trains a neural network, which connects the ensemble forecasts and possible other covariates to the distributional parameters. EMOS and DRN models for different weather quantities usually differ in the parametric family describing the predictive distribution, and the EMOS link functions and the architectures of the DRN networks might also vary. Nonparametric methods include quantile regression, providing the predictive distribution in terms of its quantiles using either statistical tools \citep{fh07,brem19} or machine learning techniques \citep{tmzn16,brem20}, or methods improving directly the raw ensemble predictions like quantile mapping \citep{hs18} or member-by-member post-processing \citep{vsv15}.

To calibrate solar irradiance ensemble forecasts, \citet{sealb21} suggest EMOS models based on censored logistic and censored normal (CN0) distributions, while \citet{bb24} and \citet{hkl24} propose DRN counterparts of the latter. Furthermore, \citet{lsbdps20} compare linear quantile regression and the analog ensemble technique to EMOS models utilizing truncated normal and truncated generalized extreme value distributions. \citet{bwks19} consider quantile regression, quantile regression neural networks, random forests, and gradient boosting decision trees, while \citet{syl24} propose a non-crossing quantile regression neural network. According to the classification of \citet{ym21}, all the post-processing methods mentioned above can be considered probabilistic-to-probabilistic (P2P) approaches, see also \citet[][Section 8.6]{yk24}.

The present work investigates probabilistic forecasts of solar irradiance for Regions III and IV in Chile, covering an area with the second-largest PV power potential after the Atacama Desert \citep{mfr17}. For this reason, eight-member ensemble forecasts of solar irradiance for the calendar year 2021 are generated using the Weather Research and Forecasting (WRF) model \citep{skamarock19}, which is then calibrated using the benchmark CN0 EMOS model and its DRN counterpart. Furthermore, we also propose a neural network-based post-processing method resulting in improved eight-member ensemble predictions. All forecasts are evaluated against station observations for 30 locations in the study area, and in a detailed case study, the skill of post-processed predictions is compared to the raw WRF ensemble. To the best of the authors' knowledge, no studies have been published yet that address the post-processing of solar irradiance ensemble forecasts for this part of the World. 

The paper is organized as follows. The description of the observed data, the WRF model configurations, and a preliminary assessment of the performance of WRF forecasts is given in Section \ref{sec2}. Section \ref{sec3} introduces the applied post-processing methods, approaches to training data selection for modelling, and the considered forecast evaluation tools. The results of our case study are reported in Section \ref{sec4}, and the paper concludes with a summary and discussion in Section \ref{sec5}.

\section{Data}
\label{sec2}

\subsection{Solar irradiance observations}
\label{subs2.1}

\begin{figure}[t]
\begin{center}
\epsfig{file=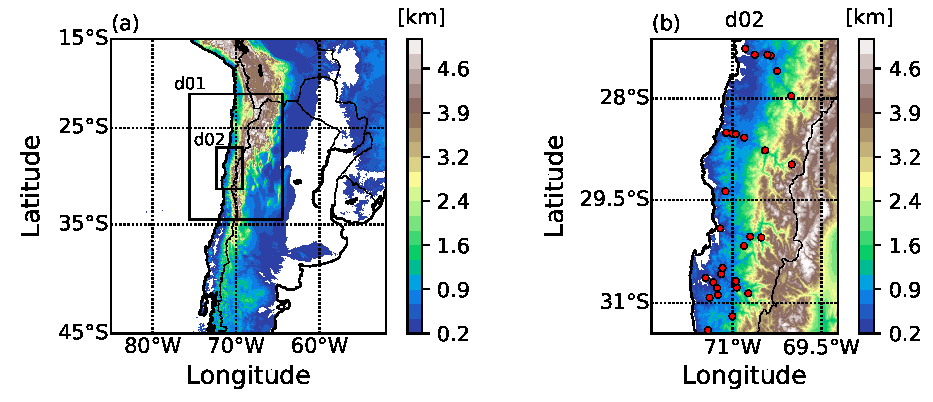, width=.95\textwidth}
\end{center}
\caption{(a) The WRF two-nested domain configuration (d01 and d02) used for each simulation and the region's topography (shaded colors), and (b) is a zoom-in to domain d02, showing the location of the 30 observation stations (red circles).}
\label{fig:wrf_domain}
\end{figure}

Solar irradiance observations (given in W/m$^2$) are provided by the National Weather Service (DMC, an acronym for the Spanish name) station network. We consider data from 30 stations in the Atacama and Coquimbo regions, between the coast and the Andes Cordillera. Table \ref{tab:radiation_stations} describes each station's name and location (latitude, longitude, altitude), and Figure \ref{fig:wrf_domain}b shows their spatial distribution over the study region. The observations with hourly temporal resolution were downloaded from the website: \url{https://climatologia.meteochile.gob.cl/}.

\begin{table}[!h]
  \caption{Description of the name, longitude, latitude, altitude, and region of 30 irradiance observation stations used in this study.}
  \begin{center}
    \begin{tabular}{rccccc}
       \hline
       No. & Name  & Longitude & Latitude & Altitude [m] & Region \\
       \hline
       1 & Desierto de Atacama, Caldera Ad.  & -70.781 & -27.254 & 197 & III\\
       2 & Universidad de Atacama  & -70.353 & -27.359 & 362 & III\\
       3 & Amolana  & -70.010 & -27.960 & 1090 & III\\
       4 & Copiap\'o  & -70.408 & -27.35 & 574 & III\\
       5 & Tierra Amarilla, T. Lautaro  & -70.000 & -28.976 & 1173 & III\\
       6 & Tierra Amarilla, Jotabeche  & -70.245 & -27.589 & 601 & III\\
       7 & La Copa  & -70.622 & -27.346 & 188 & III\\
       8 & Freirina Nicolasa  & -71.01 & -28.517 & 156 & III\\
       9 & CE Huasco  & -70.798 & -28.581 & 470 & III\\
       10 & Freirina Vallenar  & -70.943 & -28.526 & 230 & III\\
       11 & Alto del Carmen  & -70.449 & -28.768 & 822 & III\\
       12 & Freirina  & -71.104 & -28.507 & 100 & III\\
       13 & La Florida, La Serena Ad.  & -71.207 & -29.914 & 137 & IV\\
       14 & La Higuera, El Trapiche  & -71.116 & -29.372 & 281 & IV\\
       15 & Ovalle Escuela Agr\'\i cola  & -71.187 & -30.580 & 310 & IV\\
       16 & Punitaqui (FDF)  & -71.256 & -30.780 & 216 & IV\\
       17 & El Tololo  & -70.804 & -30.168 & 2154 & IV\\
       18 & Algarrobo Bajo  & -71.451 & -30.633 & 80 & IV\\
       19 & Camarico  & -71.322 & -30.699 & 290 & IV\\
       20 & El Palqui  & -70.927 & -30.774 & 504 & IV\\
       21 & Vicu\~na, Los Pimientos  & -70.697 & -30.034 & 642 & IV\\
       22 & Monte Patria, Municipalidad  & -70.958 & -30.698 & 406 & IV\\
       23 & Paiguano  & -70.511 & -30.047 & 1222 & IV\\
       24 & Ovalle Recoleta  & -71.166 & -30.491 & 403 & IV\\
       25 & Monte Patria  & -70.939 & -30.682 & 563 & IV\\
       26 & Chaguaral  & -70.731 & -30.857 & 1194 & IV\\
       27 & La Polvareda  & -71.240 & -30.883 & 265 & IV\\
       28 & Ajial de Quiles  & -71.384 & -30.920 & 464 & IV\\
       29 & Canela Baja (Liceo P. Jos\'e Herde)  & -71.413 & -31.390 & 343 & IV\\
       30 & Liceo Samuel Rom\'an Rojas  & -70.999 & -31.188 & 936 & IV\\
       \hline
    \end{tabular}
    \end{center}
    \label{tab:radiation_stations}
\end{table}

\subsection{WRF model configuration and ensemble members}
\label{subs2.2}

The Advanced Research module of the Weather Research and Forecasting (WRF) model, version 4.4.2 \citep{skamarock19}, was used in this study. We consider eight-member ensemble forecasts for solar irradiance (downward shortwave flux) for the calendar year 2021 generated by the model. All forecasts are initialized at 0000 UTC with a 1 h temporal resolution and a forecast horizon of 48 h. The eight simulations were configured with two nested domains, as depicted in Figure \ref{fig:wrf_domain}a, with 9 km (d01) and 3 km (d02) horizontal resolutions and 36 vertical levels at variable resolution and enhanced density near the surface and the tropopause. We will show only the results for the smallest domain (d02) with the highest resolution (3 km). This domain includes regions III and IV of Chile, covering several coastal and interior towns and cities to the west and part of the Andes Cordillera to the East. Forecasts from the Global Forecast System (GFS) model at $0.25^{\circ} \times 0.25^{\circ}$ horizontal resolution provided the initial and boundary conditions for the WRF model regional forecasts every three hours. The WRF simulations employed land-use data based on the Moderate Resolution Imaging Spectroradiometer (MODIS) at 15 arc seconds (approximately 0.4 km in horizontal resolution). We also set a 2-way interaction, which allows the inner domain to provide feedback to its parent domain.

The eight simulations used the Noah-MP Land Surface Model \citep{nym11,yang11} to parametrize the surface-atmosphere interaction in all domains. The Noah-MP scheme forecasts the soil temperature and moisture and provides fractional snow cover and frozen soil physics. Convective processes in all domains were calculated with the Kain–Fritsch scheme \citep{k04}, while the Thompson double moment scheme \citep{tfrh08} was used for microphysics. The eight-member simulations differ in the employed radiation and planetary boundary layer (PBL) scheme. Four radiation and five PBL schemes were combined to form the eight-member simulations, whose description is displayed in Table \ref{tab:wrfSchemes}. The Yonsei University \citep[YSU;][]{hnd06}, the Mellor-Yamada-Janji\'{c} \citep[MYJ;][]{j94}, the quasi-normal scale elimination \citep[QNSE;][]{sgp05}, the University of Washington \citep[UW;][]{bp09}, and the asymmetric convective model \citep[ACM2;][]{p07} were used to calculate the PBL processes, whereas the rapid radiative transfer model \citep[RRTMG;][]{idm08}, the Dudhia \citep{d89}, the Fu-Liou-Gu \citep[FLG;][]{glof11}, and the new Goddard scheme \citep{chou99,chou2001} were used to parametrize the longwave and shortwave radiation processes.

\begin{table}[t]
  \caption{Description of radiation (RAD), planetary boundary layer (PBL), land-surface model (LSM), cumulus, and microphysics (MP) parametrizations used on each of the eight WRF ensemble members.}
\begin{center}
  \begin{tabular}{cccccc}
    \hline
    {Member}&{PBL}&{RAD}&{LSM}&{Cumulus}&{MP}\\
    \hline
    1&YSU&RRTMG&Noah-MP&Kain–Fritsch&Thompson\\
    2&YSU&Dudhia&Noah-MP&Kain–Fritsch&Thompson\\
    3&YSU&FLG&Noah-MP&Kain–Fritsch&Thompson\\
    4&YSU&Goddard&Noah-MP&Kain–Fritsch&Thompson\\
    5&MYJ&RRTMG&Noah-MP&Kain–Fritsch&Thompson\\
    6&QNSE&RRTMG&Noah-MP&Kain–Fritsch&Thompson\\
    7&UW&RRTMG&Noah-MP&Kain–Fritsch&Thompson\\
    8&ACM2&RRTMG&Noah-MP&Kain–Fritsch&Thompson\\
    \hline
\end{tabular}
\end{center}
\label{tab:wrfSchemes}
\end{table}

\subsection{Forecast skill of the WRF ensemble}
\label{subs2.3}

\begin{figure}[t]
\begin{center}
\epsfig{file=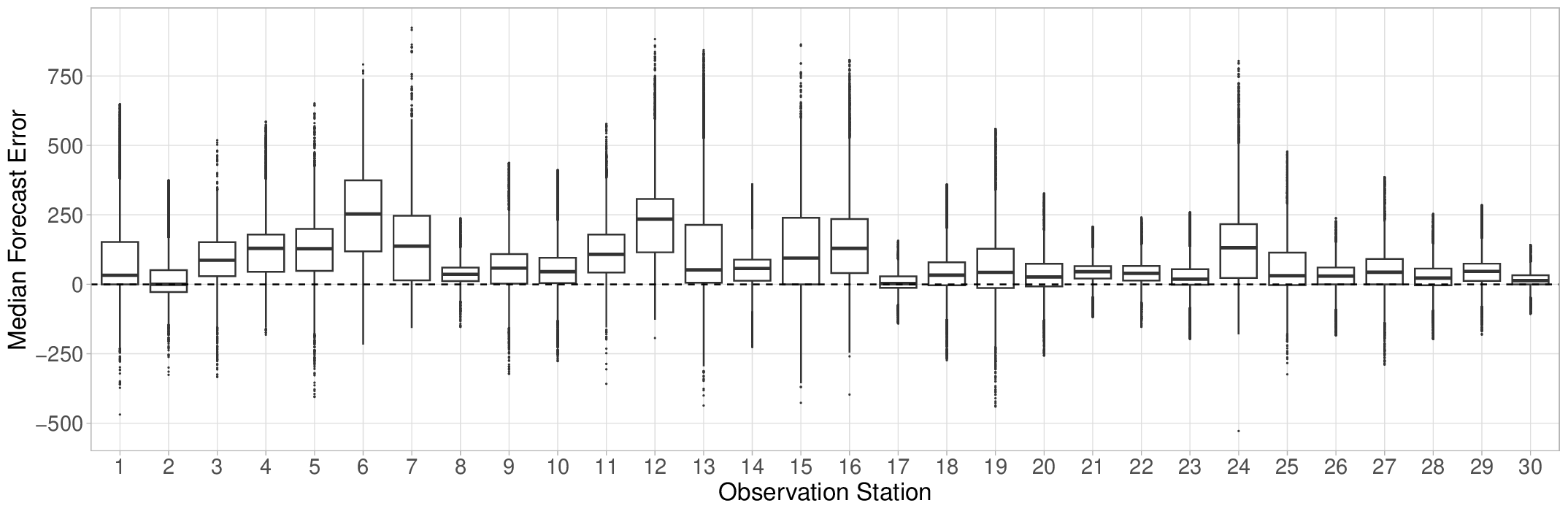, width=\textwidth}
\end{center}
\caption{Median forecast error (W/m$^2$) of the WRF ensemble at the observation stations listed in Table \ref{tab:radiation_stations} for all dates and lead times 12 -- 24 h and 36 -- 48 h.}
\label{fig:medForcErrSt}
\end{figure}

\begin{figure}[t]
\begin{center}
\epsfig{file=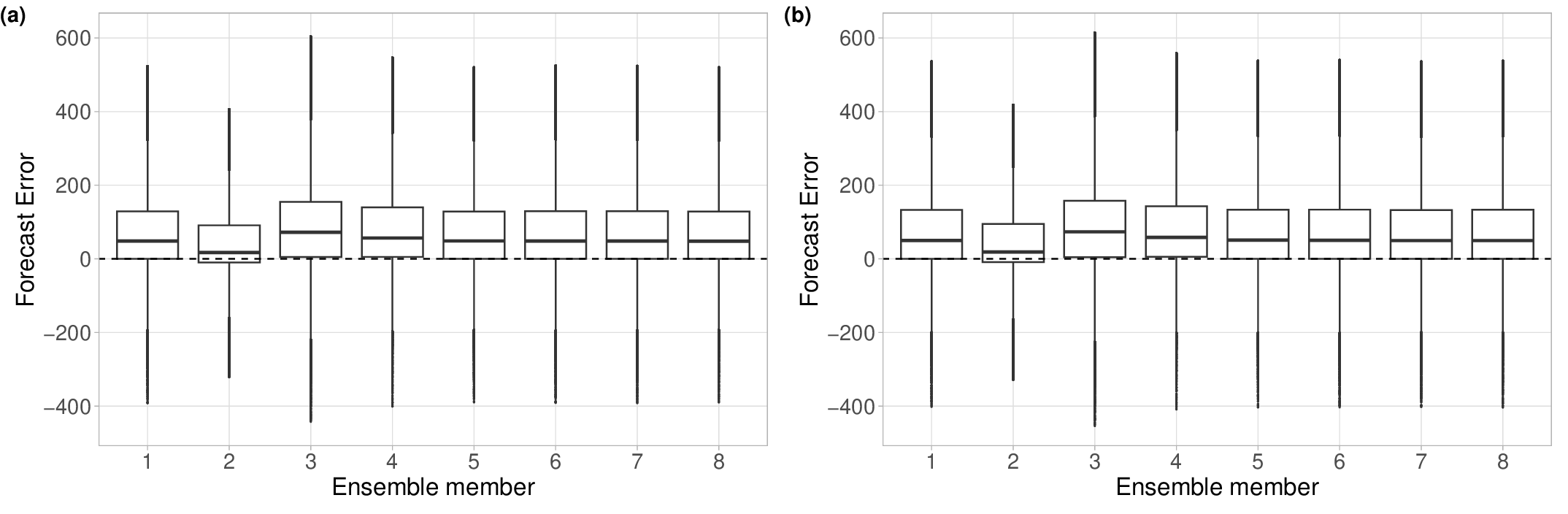, width=\textwidth}
\end{center}
\caption{Forecast error (W/m$^2$) of the individual ensemble members for all dates and locations for lead times (a) 12 -- 24 h and (b) 36 -- 48 h.}
\label{fig:forcErrEns}
\end{figure}

\begin{figure}[t]
\begin{center}
\epsfig{file=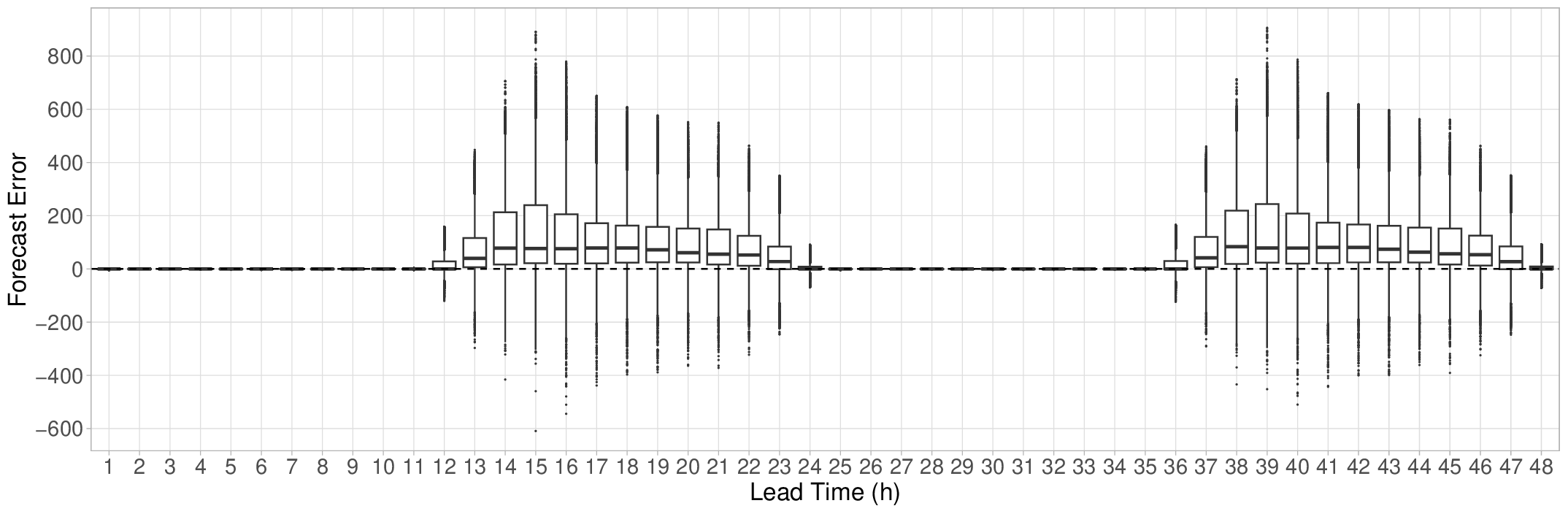, width=\textwidth}
\end{center}
\caption{Forecast error (W/m$^2$) of the ensemble median for all dates and locations as function of the lead time.}
\label{fig:errEnsMed}
\end{figure}

The matching of the WRF ensemble predictions with observations at monitoring stations listed in Table \ref{tab:radiation_stations} is performed by extracting forecasts for the nearest grid points from the WRF domain d02. As depicted in Figure \ref{fig:wrf_domain}b, the topography of this region is rather complex; the station altitudes range from 80 m to 2154 m, which variability has an impact on the forecast performance. Figure \ref{fig:medForcErrSt} shows the box plots of the median forecast error of the WRF ensemble at the various observation stations for the whole calendar year 2021 for lead times corresponding to the period between 1200 and 2400 UTC when positive irradiance is likely to be observed. Raw WRF forecasts systematically overestimate the actual irradiance and both the magnitude of the bias and the spread of the median forecast error strongly depend on the location. However, in contrast to \citet{dnmb20}, where the station altitude played a key role in the capability of WRF temperature forecasts, here, despite the WRF irradiance ensemble performing the best at the highest station (No. 17; El Tololo), one cannot find a clear connection between the elevation and the forecast skill. The same positive bias can be observed in Figure \ref{fig:forcErrEns}, displaying the box plots of the forecast error of the individual ensemble members for all dates and locations, treating separately shorter (12 -- 24 h) and longer (36 -- 48 h) lead times. Although longer forecast horizons result in slightly larger forecast errors, Figure \ref{fig:forcErrEns}a and Figure \ref{fig:forcErrEns}b convey the same message. Ensemble member 2 substantially outperforms the other 7 members; the largest error corresponds to member 3, followed by member 4, whereas the performance of members 1 and 5 -- 6 is fairly similar. Finally, the box plots of the diurnal evolution of the forecast error of the ensemble median depicted in Figure \ref{fig:errEnsMed} indicate that the positive bias is systematic also along the forecast horizons, and the largest errors correspond to 1500 and 1600 UTC when the solar irradiance reaches its peak.

\section{Post-processing and forecast evaluation methods}
\label{sec3}

As mentioned in the Introduction, we consider two different parametric methods for post-processing WRF solar irradiance ensemble forecasts. The key step in parametric modelling is the choice of the predictive distribution. Addressing non-negativity of solar irradiance, several parametric models utilize distributions left truncated from below at zero, see e.g. \citet{lsbdps20} or \citet{yang20}. However, such models require the specification of times of day with positive irradiance, whose periods strongly depend on the location and season of the year. This deficiency can be solved by considering laws that assign a positive mass to the event of zero irradiance, as such a distribution can handle even the night hours when both predicted and observed irradiance are zero. A popular choice is to left censor a suitable distribution at zero, which approach proved to be successful e.g. in modelling precipitation accumulation \citep{sch14,bn16}. Here we utilize the censored normal EMOS approach of \citet{sealb21}, considered as a benchmark method, and the corresponding DRN model investigated by \citet{bb24} and \citet{hkl24}.

Beyond parametric modelling,  we also suggest a neural network-based nonparametric technique, where the output is a calibrated forecast ensemble of the same cardinality as the original raw one.

In what follows, let \ $f_1, f_2, \ldots ,f_8$ \ denote the eight-member WRF irradiance forecast of a given forecast horizon for a given location and time, and denote by \ $\overline f$ \ and \ $S^2$ \ the corresponding ensemble mean and variance, respectively.

\subsection{EMOS model for solar irradiance}
\label{subs3.1}
Let \ $G(x|\mu,\sigma)$ \ denote the cumulative distribution function (CDF) of a Gaussian distribution with mean \ $\mu$ \ and standard deviation \ $\sigma >0$, \ that is \
\begin{equation*}
G(x|\mu,\sigma):=\Phi \left(\frac{x-\mu}\sigma \right), \qquad x\in{\mathbb R},
\end{equation*}
with \ $\Phi$ \ denoting the CDF of a standard normal law. Then the CDF of a normal distribution with location \ $\mu$ \ and scale \ $\sigma$ \ left-censored at zero (CN0) equals
\begin{equation}
 \label{eq:cn0CDF}
  G_0^c(x|\mu,\sigma):=\begin{cases} G(x|\mu,\sigma),& \quad x\geq 0, \\
    0, & \quad x<0. \end{cases}
\end{equation}
This distribution assigns mass \ $G_0^c(0|\mu,\sigma)$ \ to the origin and has mean
\begin{equation*}
  \kappa = \mu \Phi\big(\mu/\sigma\big) + \sigma \varphi\big(\mu/\sigma\big),
\end{equation*}
where \ $\varphi$ \ denotes the standard normal probability density function (PDF). Following \citet{sealb21} and \citet{bb24}, the ensemble members are linked to the parameters of the CN0 distribution via equations
\begin{equation*}
  \label{eq:link}
  \mu = \gamma_0 + \gamma_1 \overline f + \gamma_2p_0 \qquad \text{and} \qquad \sigma = \exp \big(\delta _0 + \delta_1 \log S\big),
\end{equation*}
where \ $p$ \ is the proportion of WRF ensemble members predicting zero irradiance, that is
\begin{equation*}
  p_0 := \frac 18 \sum_{k=1}^8 {\mathbb I}_{\{f_k=0\}}
\end{equation*}
with \ ${\mathbb I}_H$ \ denoting the indicator function of a set \ $H$. \  Model parameters \ $\gamma_0,\gamma_1,\gamma_2,\delta_0,\delta_1 \in {\mathbb R}$ \ are estimated following the optimum score principle of \citet{gr07}, that is obtained by minimizing the mean value of a proper scoring rule  (in our case the continuous ranked probability score (CRPS) defined by \eqref{eq:CRPSdef} in  Section \ref{subs3.5}) over appropriate training data comprising past forecast-observation pairs.

\subsection{DRN model for solar irradiance}
\label{subs3.2}
Distributional regression networks (DRNs), first employed by \citet{rl18} for calibrating temperature ensemble forecasts, represent an advanced class of machine learning-based post-processing models. Similar to the EMOS approach, they extend traditional regression techniques by predicting parameters of the forecast distribution belonging to a given family; in our case, the location \ $\mu$ \ and scale \ $\sigma$ \ of the CN0 distribution specified by CDF \eqref{eq:cn0CDF}. The DRN approach enhances the statistical calibration of ensemble forecasts, providing a more comprehensive understanding of prediction uncertainty. DRNs typically leverage predictor variables, such as NWP quantities and station characteristics, to inform their predictions. Additionally, station embeddings enable the network to recognize location-specific information, capturing unique features and patterns relevant to individual stations.

DRNs are often implemented using a multilayer perceptron \citep[MLP;][]{dlbook} architecture, which improves their ability to capture complex relationships between the NWP forecasts and other covariates and the distributional parameters to be predicted. The feedforward structure of the network allows input features to pass through multiple hidden layers, where neurons apply weighted transformations and activation functions to the input signals. Activation functions such as ReLU, sigmoid, or tanh introduce non-linearity, enabling the model to learn intricate patterns and relationships in the data. Similar to EMOS modelling, the weights of the MLP neural networks are estimated by minimizing the mean CRPS of the CN0 predictive distributions over the training data, optimizing probabilistic forecasts to align with observed outcomes. Note that building upon the work of \citet{rl18}, which focused on fully connected networks such as MLP, recently convolutional layers have also gained popularity within DRNs \citep{vwds21, lpxd22}. These layers leverage local patterns and spatial hierarchies in the data, further enhancing the model’s ability to capture complex relationships. However, in this study, we keep the traditional MLP architecture.

To further enhance model performance and prevent overfitting, early stopping is often employed as a regularization technique during training. By monitoring performance on a validation dataset, training halts when improvements cease, ensuring the model retains its generalization capabilities. This approach not only optimizes the training process but also contributes to more robust predictions.

\subsection{Machine learning-based forecast improvement }
\label{subs3.3}
In addition to constructing predictive distributions, another possible alternative is statistical post-processing using machine learning models that generate improved ensemble forecasts. For this purpose, similar to the DRN model described in Section \ref{subs3.2}, we use a neural network based on the MLP architecture; however, here, the number of neurons in the output layer equals the number of ensemble members to be generated. The other principal difference is the implementation of the loss function.  Due to the nature of the output, here we apply the ensemble CRPS given by \eqref{eq:ensCRPSdef} (see Section \ref{subs3.5}), with the constraint that the predicted solar irradiance forecasts can only be non-negative. This post-processing method is flexible as it is distribution-free, and the number of ensemble members to be generated is up to the user, provided enough training data is available. However, to ensure direct comparability with the WRF forecasts, we create an 8-member corrected ensemble for each verification day, location, and forecast horizon. 

\subsection{Training data selection}
\label{subs3.4}
The efficiency of all post-processing methods, including the ones described in Sections \ref{subs3.1} -- \ref{subs3.3}, strongly depends on the spatial and temporal decomposition of the training data. 

From the point of view of temporal selection, a popular approach is using a sliding window, where the model for a given date is trained using forecast-observation pairs from the preceding \ $n$ \ calendar days. This simple method, also utilized in our study, allows a quick adaptation to seasonal variations or model changes; nevertheless, larger time shifts such as monthly, seasonal, or yearly windows might also be beneficial \citep[see e.g.][]{jmg23}. Alternatively, one can consider a fixed, very long training period, a popular approach in machine learning-based post-processing methods requiring a large amount of training data \citep[see e.g.][]{sl22,hkl24}. For a systematic comparison of time-adaptive model training schemes, we refer to \citet{llmssz20}.

Regarding the spatial composition of training data, the traditional approaches are local and regional (global)  modelling \citep{tg10}. Local models are based on past forecast-observation pairs for the actual location under consideration, whereas the regional approach utilizes historical data of all studied stations. In general, local models outperform their regional counterparts, provided one has enough location-specific training data to avoid numerical issues during the calibration process. The regional approach, resulting in a single set of EMOS parameters or neural network weights for the whole ensemble domain, is more suitable if only short training periods are allowed; however, it might hardly handle large and heterogeneous areas. To combine the advantageous properties of the above two approaches to spatial selection, \citet{lb17} suggest a novel clustering-based semi-local method which appeared to be successful for several different weather variables and ensemble domains, see e.g. \citet{bbpbb20}, \citet{szgb23}, or \citet{bl24}. For a given date of the verification period, a feature vector is assigned to each observation station, representing the station climatology and the forecast error of the ensemble mean during the training period. Based on these feature vectors, the stations are then grouped into clusters using $k$-means clustering \citep[see e.g.][Section 16.3.1]{w19}, and within each cluster, regional modelling is performed. In the case of rolling training windows, the stations are regrouped dynamically for each particular training set.

\subsection{Forecast verification}
\label{subs3.5}
As mentioned, the parameters of the CN0 EMOS model described in Section \ref{subs3.1} are estimated by minimizing the mean of a proper scoring rule over the training data. In particular, we consider the continuously ranked probability score \citep[CRPS;][Section 9.5.1]{w19}, which is probably the most popular proper verification score in atmospheric sciences. Given a probabilistic forecast expressed as a predictive CDF \ $F$ \ and an observation \ $x\in {\mathbb R}$, \ the CRPS is defined as
\begin{equation}
    \label{eq:CRPSdef}
\crps(F,x) := \int_{-\infty}^{\infty}\Big[F(y)-{\mathbb I}_{\{y\geq x\}}\Big]^2{\mathrm d}y ={\mathsf E}|X-x|-\frac 12
{\mathsf E}|X-X'|,
\end{equation}
where  \ $X$ \ and \ $X'$ \ are independent random variables distributed according to \ $F$ \ and having a finite first moment. The CRPS is a negatively oriented score (smaller values indicate better performance) and it simultaneously assesses both the calibration and the sharpness of a probabilistic forecast. Calibration means a statistical consistency between forecasts and observations, while sharpness refers to the concentration of the probabilistic forecasts. Furthermore, the expression of the CRPS on the right-hand side of \eqref{eq:CRPSdef} indicates that it can be expressed in the same unit as the observation. For the censored normal distribution considered in Sections \ref{subs3.1} and \ref{subs3.2}, the CRPS has a closed form \citep[see e.g.][]{jkl19}, which allows a computationally efficient numerical optimization, hence making it eligible to serve as a loss function both in EMOS and DRN modelling. For a forecast ensemble \ $f_1,f_2, \ldots ,f_K$, \ one should consider the empirical CDF \ $\widehat F_K$, \ resulting in the expression 
\begin{equation}
  \label{eq:ensCRPSdef}
\crps(\widehat F_K,x)=\frac 1K\sum_{k=1}^K\big |f_k -x \big| - \frac 1{2K^2} \sum_{k=1}^K\sum_{\ell =1}^K\big | f_k - f_{\ell}\big |, 
\end{equation}
see e.g. \citet{kltg21}. The same definition applies to the corrected ensemble forecast produced by the approach described in Section \ref{subs3.3} and calibrated samples generated from the EMOS or DRN predictive distributions. Note that the above formula slightly differs from the ensemble CRPS given in \citet[][Section 9.7.3]{w19}; however, in the  {\tt R} package {\tt scoringRules} \citep{jkl19}, expression \eqref{eq:ensCRPSdef} is implemented.

In the case study of Section \ref{sec4}, the predictive performance of the competing probabilistic forecasts with a given forecast horizon is compared with the help of the mean CRPS over all forecast cases in the verification period. Furthermore, we also quantify the improvement in the mean CRPS of a probabilistic forecast \ $F$ \ for a reference forecast \ $F_{\text{ref}}$ \ using the continuous ranked probability skill score \citep[CRPSS; see e.g.][]{gr07}. This positively oriented quantity (the larger, the better) is defined as
  \begin{equation*}
    \label{eq:CRPSSdef}
   \crpss := 1 - \frac{\overline\crps_F}{\overline\crps_{F_{\text{ref}}}},
 \end{equation*}
where  \ $\overline\crps_F$ \ and \ $\overline\crps_{F_{\text{ref}}}$ \ denote the mean CRPS corresponding to forecasts \ $F$ \ and \ $F_{\text{ref}}$, \ respectively.

A separate assessment of calibration and sharpness of predictive distributions can be obtained with the help of the coverage and average width of \ $(1-\alpha)100\,\%, \ \alpha \in ]0,1[,$ \ central prediction intervals, respectively. By coverage, we mean the proportion of observations located between the \ $\alpha/2$ \ and \ $1-\alpha/2$ \ quantiles of the predictive CDF, where properly calibrated forecasts result in values around \ $(1-\alpha)100\,\%$ \citep[see e.g.][]{gr07}. For a $K$-member forecast ensemble, one usually considers the nominal coverage \ $(K-1)/(K+1)100\,\%$ \ of the ensemble range ($77.78\,\%$ \ for the 8-member WRF ensemble at hand), which is the probability of the event that the rank of the observation for a calibrated prediction is greater than $1$ and less than \ $K+1$. \ Choosing \ $\alpha$ \ to match this nominal coverage allows a fair comparison of ensemble forecasts with forecasts provided as full predictive distributions.

A further plausible tool for evaluating the calibration of ensemble forecasts is the verification rank histogram (or Talagrand diagram). The Talagram diagram displays the ranks of the verifying observation with respect to the corresponding ensemble prediction \citep[][Section 9.7.1]{w19}, which for a calibrated $K$-member ensemble should follow a discrete uniform law on \ $\{1,2, \ldots ,K+1\}$. \ The shape of a rank histogram reflects the source of the lack of calibration: $\cup$- and $\cap$-shaped histograms refer to under- and overdispersion, respectively, whereas biased forecasts result in triangular shapes. Furthermore, the deviation of the distribution of the verification ranks from the uniform law can be quantified with the help of the reliability index:
\begin{equation*}
   \label{eq:relind}
 \ri:=\sum_{r=1}^{K+1}\Big| \rho_r-\frac 1{K+1}\Big|,
\end{equation*}
where \ $\rho_r$ \ denotes the relative frequency of rank \ $r$ \ over all forecast cases in the verification period \citep{dmhzds06}. The continuous counterpart of the Talagrand diagram is the probability integral transform \citep[PIT;][Section 9.7.1]{w19} histogram. The PIT is defined as the value of predictive CDF at the verifying observation with possible randomization at the points of discontinuity. For a calibrated predictive distribution, PIT is standard uniform, and the interpretation of the various deviations of shapes  of the PIT histograms from uniformity is similar to the Talagrand diagrams.

Furthermore, the accuracy of point forecasts such as forecast median is evaluated by the mean absolute error (MAE), as the median minimizes this score  \citep{gneiting11}.

Finally, the statistical significance in score differences is assessed by accompanying some skill scores and score differences by \ $95\,\%$ block bootstrap confidence intervals. We consider 2000 samples calculated using the stationary bootstrap scheme with block lengths following a geometric distribution with a mean proportional to the cube root of the length of the investigated time interval \citep{pr94}.

\subsection{Modelling and implementation details}
\label{subs3.6}
In the case of the EMOS modelling, all 48 forecast horizons are treated separately, and model parameters are estimated using a clustering-based semi-local approach. Similar to \citet{lb17}, we consider 24-dimensional feature vectors, where half of the features comprise equidistant quantiles of the climatological CDF over the training period, and the other half consists of equidistant quantiles of the empirical CDF of the forecast error of the ensemble mean. After testing several combinations of the training period length and the number of clusters, an 85-day rolling window is chosen, and in general, six clusters are formed (an average of 425 forecast-observation pairs for each estimation task), provided each cluster contains at least three observation stations. Otherwise, the number of clusters is reduced, which might result in regional modelling.

Machine learning-based forecasts are estimated regionally, and a single MLP neural network is trained for all lead times. CN0 DRN models use 20-day rolling training windows, while corrected forecasts depend on forecast-observation pairs of the preceding 25 days. These training period lengths are the results of detailed data analysis.

Through testing various hidden layer configurations for the CN0 DRN approach, we identified an optimal MLP model with two hidden layers, each containing 255 neurons, and considered a batch size of 1200. To enhance numerical stability, we standardize input features and remove missing data. The model is trained with the ``Adam'' optimizer, a learning rate of 0.01, and the ReLU activation function. Input features include the ensemble mean \ $\overline f$ \ and variance \ $S^2$, \ the proportion \ $p_0$ \ of zero-irradiance forecasts, station coordinates (latitude, longitude, altitude), and lead times. Feature importance was also assessed to evaluate each input’s impact on model performance. The model output directly represents the parameters \ $\mu$ \ and \ $\sigma$ \ of the CN0 predictive distribution, with only a squaring applied to the scale parameter to ensure non-negativity.

The training is capped at 500 epochs, but an early stopping callback -- using 20\,\% of the training data as a validation set -- often enabled convergence at approximately 50 epochs for the first verification day, with fewer epochs required on subsequent days. To address the randomness inherent in training, for each forecast case, we train the network ten times, deriving final predictions by averaging the distribution parameters across these sessions.

Our detailed tests suggest that when DRN input features are restricted to ensemble statistics used in EMOS modeling -- i.e., the same set of explanatory variables -- the DRN's performance generally aligns with that of EMOS. This indicates that without additional input information beyond the functionals of the ensemble forecasts, DRNs may have limited capacity for performance gains over EMOS in terms of accuracy and calibration. Notably, including station location information significantly contributed to enhanced model performance, which aligns with the findings of \citet{hkl24}.  
  
For the MLP network providing corrected forecasts, we also comprehensively tested different input features and hyperparameters, though optimal results were achieved with the same settings as for the CN0 DRN approach. Due to the increase of the neurons in the output layer from two to eight, the number of weights to be estimated increased from 67,832 to 69,368, which explains the 5-day increase in the training period length. Furthermore, to ensure the non-negativity of the generated forecasts, we take the maximum of zero and the predicted value, which is consistent with the handling of negative predictions in the loss function. Finally, as before, to account for the stochastic nature of the model, ten independent runs are performed for each forecast case. The generated forecasts are then sorted in ascending order, and the average is computed over the sorted values.

\section{Results}
\label{sec4}
In the following case study, the predictive performance of the CN0 EMOS and CN0 DRN approaches described in Sections \ref{subs3.1} and \ref{subs3.2}, respectively, is evaluated together with the machine learning-based forecast improvement method presented in Section \ref{subs3.3}. To ensure a fair comparison of the latter with the EMOS and DRN models resulting in full predictive distributions, we also investigate the forecast skill of 8-member samples generated from the corresponding CN0 distributions, where one can consider their \ $1/9, \ 2/9, \ldots , 8/9$ \ quantiles, simple random sampling, or stratified samples \citep{hsa16}. As preliminary analysis reveals only minor differences in the skill of equidistant quantiles and stratified samples, both outperforming random sampling, here we work with equidistant quantiles of CN0 EMOS and CN0 DRN predictive distributions. The corresponding forecasts are referred to as CN0 EMOS-Q and CN0 DRN-Q, respectively.

In the following analysis, the predictive performance of the competing post-processing methods is tested on forecast-observation pairs for the nine months of 1 April -- 31 December 2021 (275 calendar days, right after the first 85-day training window used in EMOS modelling), and the raw WRF irradiance forecasts are used as a reference.

\begin{figure}[t]
\begin{center}
\epsfig{file=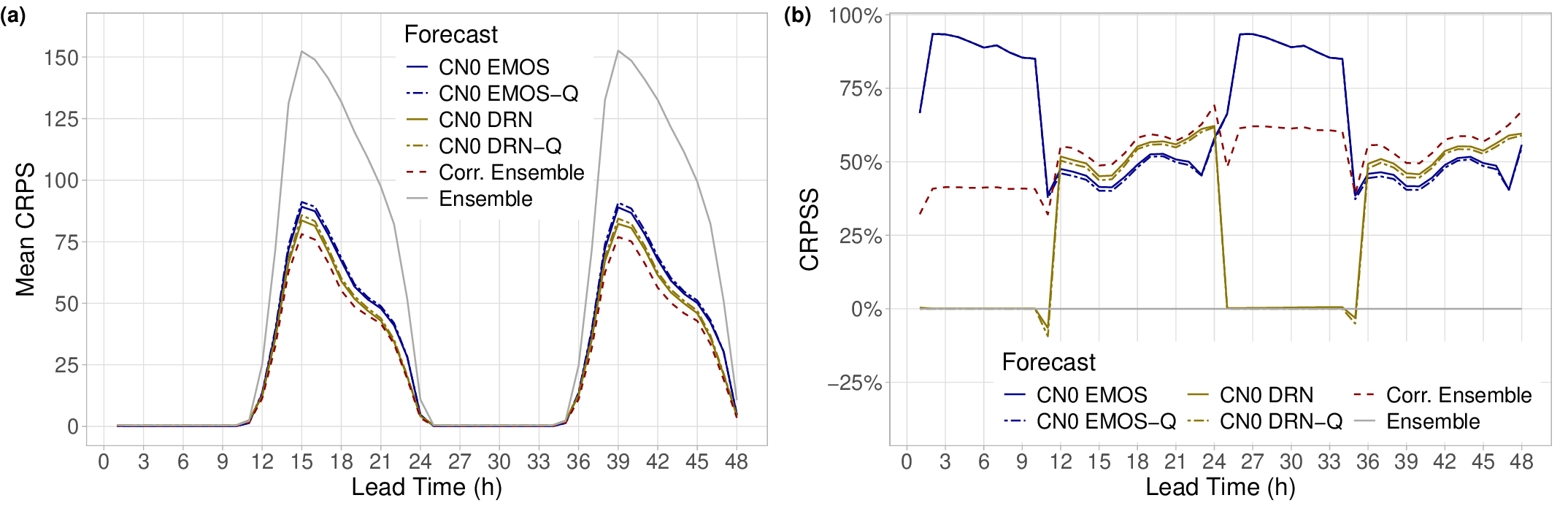, width=\textwidth}
\end{center}
\caption{(a) Mean CRPS of raw and post-processed irradiance forecasts and (b) CRPSS of post-processed forecasts with respect to the raw WRF ensemble as functions of the lead time.}
\label{fig:crps_crpss}
\end{figure}

\begin{table}[t]
  \caption{Overall mean CRPS of post-processed irradiance forecasts as a proportion of the mean CRPS of the raw WRF ensemble for observed irradiance not less than 7.5\, W/m$^2$.}
  \begin{center}
\begin{tabular}{c|c|c|c|c}
  CN0 EMOS&CN0 EMOS-Q&CN0 DRN&CN0 DRN-Q&Corr. Ensemble\\ \hline
  52.58\,\%&53.69\,\%&47.90\,\%&49.04\,\%&44.57\,\%
\end{tabular}
\end{center}
\label{tab:crps}
\end{table}

Figure \ref{fig:crps_crpss}a displays the mean CRPS of raw and post-processed WRF irradiance forecasts over all 30 locations and 275 days in the verification period. Between 1200 and 2400 UTC (lead times 12 -- 24 h and 36 -- 48 h), when positive irradiance is likely to be observed,
all calibration methods outperform the raw ensemble by a wide margin. According to the CRPSS values of Figure \ref{fig:crps_crpss}b, at this time of the day, the improvement with respect to the raw ensemble is around 50\,\% for all competing forecasts, and the ranking of the various methods is consistent for all forecast horizons. 
There are only minor differences in skill between forecasts resulting in full predictive distributions (CN0 EMOS and CN0 DRN) and their sample counterparts (CN0 EMOS-Q and CN0 DRN-Q, respectively). The corrected ensemble results in the highest CRPSS, followed by the DRN and EMOS predictions. Note that for 1 -- 11 h and 25 -- 35 h forecasts, the skill score values are irrelevant; they are just results of numerical issues.

\begin{figure}[t]
\begin{center}
\epsfig{file=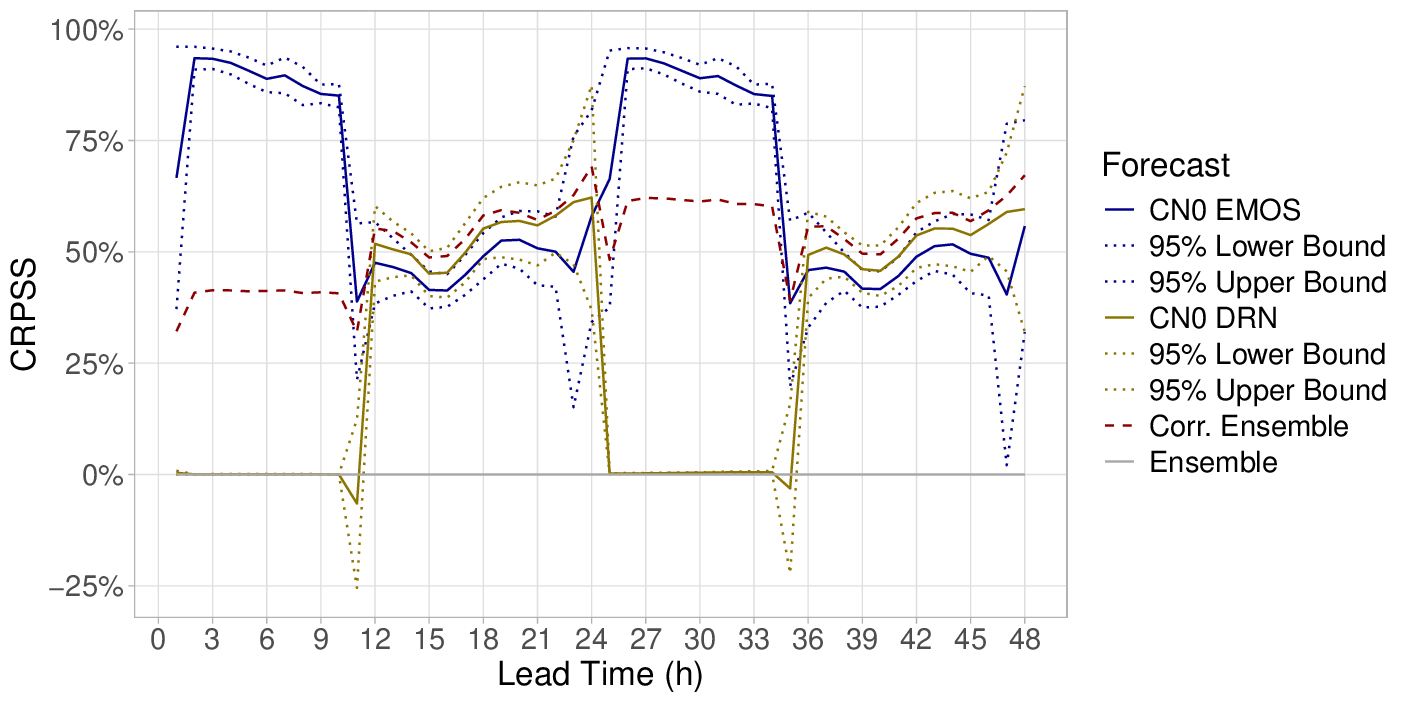, width=.66\textwidth}
\end{center}
\caption{CRPSS with respect to the raw WRF ensemble of the CN0 EMOS and CN0 DRN models (together with 95\,\% confidence intervals) and the corrected ensemble as functions of the lead time.}
\label{fig:boot_crpss}
\end{figure}

The clear ranking of the various calibrated forecasts is also confirmed by Table \ref{tab:crps}, summarizing their mean CRPS as a proportion of the mean CRPS of the WRF ensemble for forecast cases when the observed irradiance is at least 7.5\, W/m$^2$. Note that from the point of view of PV energy production, only these cases are of any interest, and the threshold coincides with the one considered in \citet{bb24} and suggested by the forecasters of the Hungarian Meteorological Service.

To investigate the statistical significance of the differences in terms of the mean CRPS, in Figure \ref{fig:boot_crpss}, the corresponding 95\,\% confidence intervals are added to the CRPSS values of the CN0 EMOS and CN0 DRN approaches. For lead times corresponding to the 1200 -- 2400 UTC interval, neither the difference between the corrected ensemble and the CN0 DRN nor the deviation of the CN0 EMOS and the CN0 DRN is significant; however, between 13 -- 20 h and 37 -- 44 h, the best-performing corrected ensemble significantly outperforms the worst-performing CN0 EMOS.

\begin{figure}[t]
\begin{center}
\epsfig{file=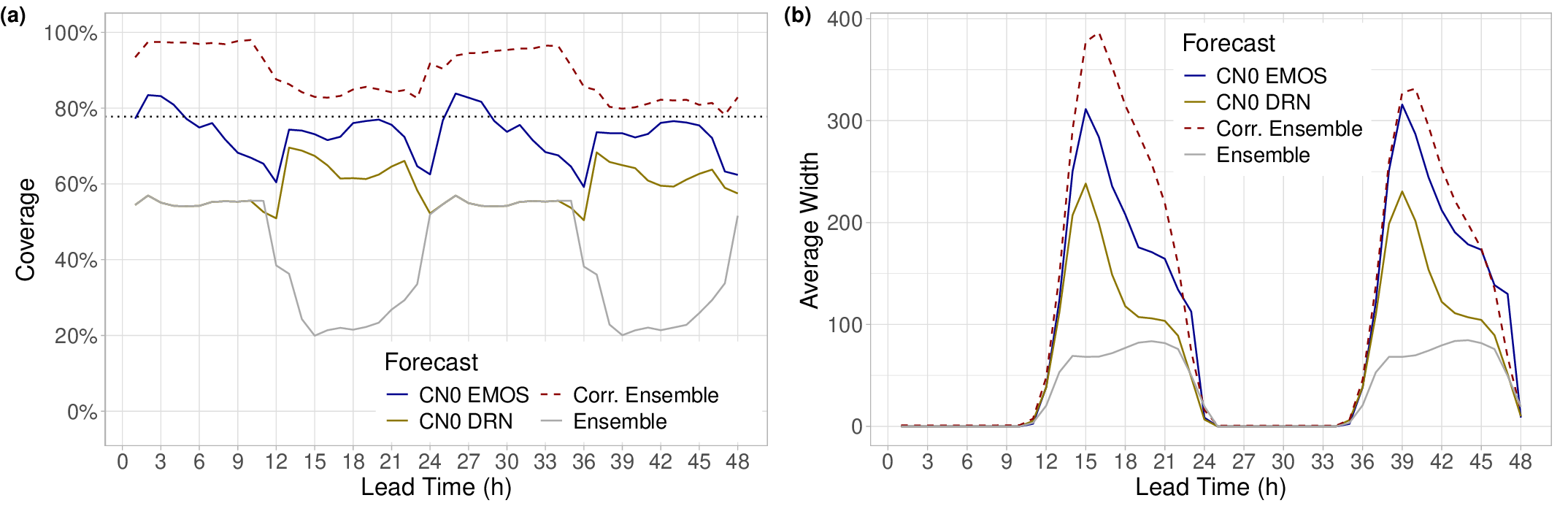, width=\textwidth}
\end{center}
\caption{(a) coverage and (b) average width of the nominal 77.78\,\% central prediction intervals of post-processed and raw irradiance forecasts as functions of the lead time. In (a), the ideal coverage is indicated by the horizontal dotted line.}
\label{fig:cov_aw}
\end{figure}

The improved calibration of post-processed forecasts can also be observed in Figure \ref{fig:cov_aw}a, displaying the coverage of the nominal 77.78\,\% central prediction intervals. To ensure comparability with the raw ensemble,  for calibrated 8-member forecasts, we consider the coverage of the whole ensemble range. In the case of CN0 EMOS-Q and CN0 DRN-Q predictions based on equidistant quantiles of the corresponding predictive distributions, the ensemble ranges coincide with the 77.78\,\% central prediction intervals of the CN0 EMOS and CN0 DRN models, respectively, so they are excluded from the analysis. At the hours of the highest solar irradiance (1300 -- 2200 UTC), the coverage of the raw WRF forecasts is below 40\,\%, where post-processed forecasts result in coverage values much closer to the targeted 77.78\,\%. A possible ranking of the three investigated methods can be formed with the help of Table \ref{tab:cov}, providing the mean absolute deviation in coverage from the nominal level for 1300 -- 2200 UTC observations. According to Figure \ref{fig:cov_aw}b, the improved coverage of calibrated forecasts is accompanied by wider prediction intervals; however, the CN0 EMOS prediction resulting in the best coverage values is considerably sharper than the second-best corrected ensemble.

\begin{figure}[t]
\begin{center}
\epsfig{file=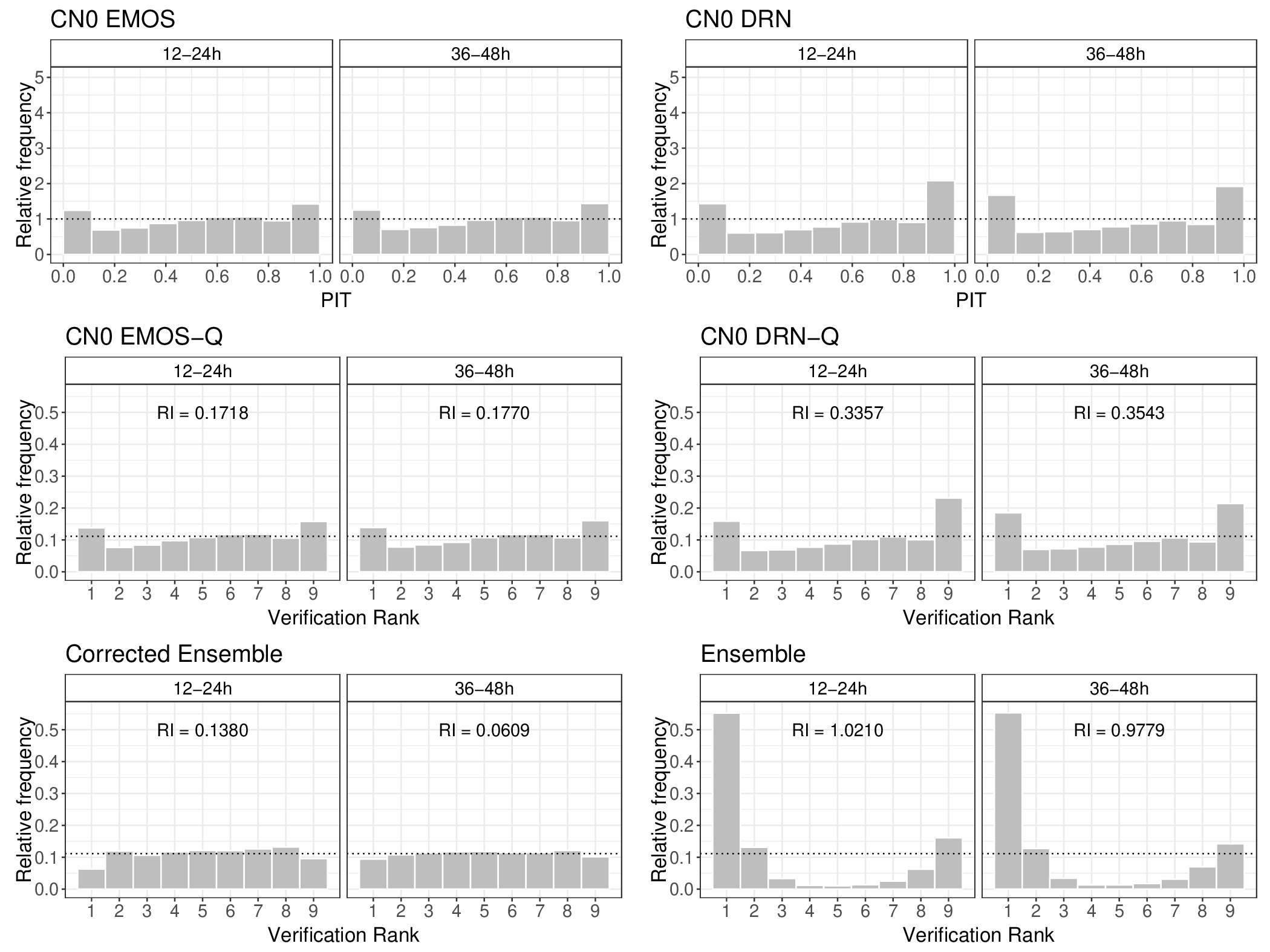, width=\textwidth}
\end{center}
\caption{PIT histograms of CN0 EMOS and CN0 DRN predictive distributions and rank histograms of raw and post-processed eight-member irradiance forecasts together with the corresponding reliability indices for lead times 12--24 h and 36--48 h.}
\label{fig:rankhist}
\end{figure}

The verification rank and PIT histograms of Figure \ref{fig:rankhist}, where we again consider only lead times corresponding to 1200 -- 2400 UTC observations, also confirm the positive effect of statistical post-processing. First, note that the PIT histograms of the CN0 predictive distributions and the rank histograms of the corresponding samples are almost identical, which agrees with the matching CRPS and CRPSS values of Figure \ref{fig:crps_crpss}. The raw WRF forecasts indicate strong underdispersion and a pronounced positive bias, resulting in reliability indices of $1.0210$ (12 -- 24 h forecasts) and $0.9779$ (36 -- 48 h forecasts). These deficiencies are substantially reduced after calibration, implying a strong improvement in reliability indices. CN0 EMOS-Q and CN0 DRN-Q forecasts are still slightly underdispersive and biased, showing reliability indices of $0.1718, \ 0.1770$ (CN0 EMOS-Q) and $0.3357, \ 0.3543$ (CN0 DRN-Q), whereas the corrected ensemble is a bit overdispersive, especially for short lead times. However, this forecast results in the lowest reliability indices of $0.1380$ and $0.0609$. 

\begin{table}[t]
  \caption{Mean absolute deviation in coverage from the nominal 77.78\,\% level over lead times corresponding to 1300 -- 2200 UTC observations.}
  \begin{center}
\begin{tabular}{c|c|c|c}
  CN0 EMOS&CN0 DRN&Corr. Ensemble&Ensemble\\ \hline
   3.49\,\%&13.84\,\%&5.18\,\%&53.22\,\%  
\end{tabular}
\end{center}
\label{tab:cov}
\end{table}

\begin{figure}[t]
\begin{center}
\epsfig{file=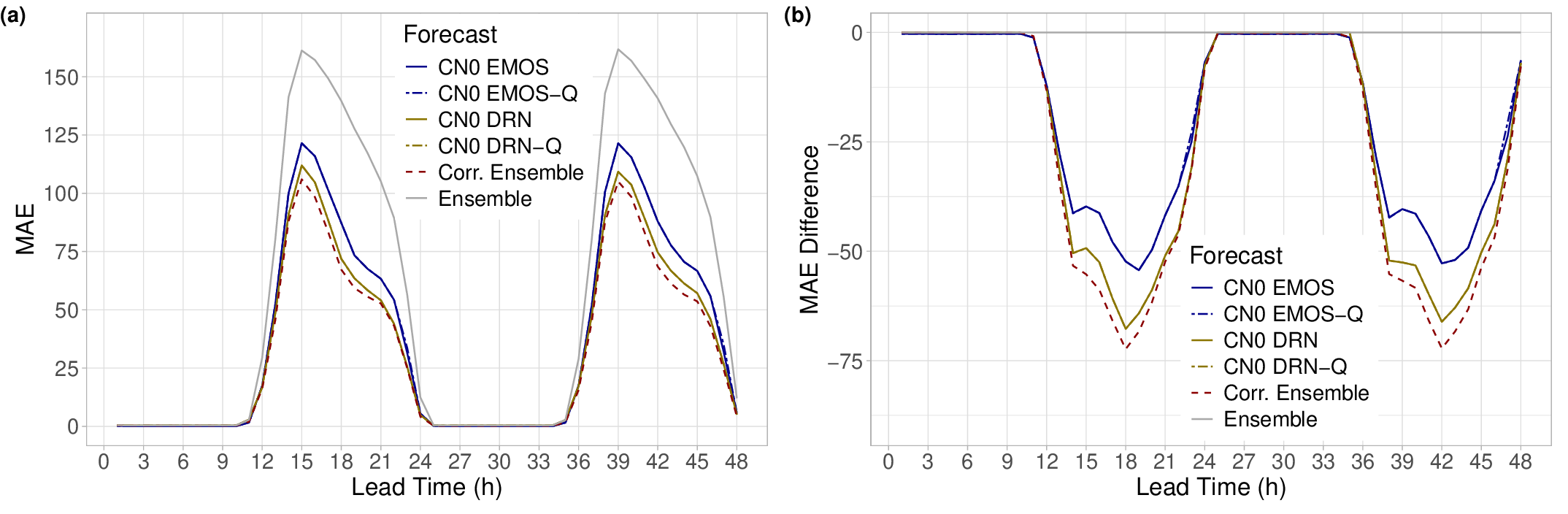, width=\textwidth}
\end{center}
\caption{(a) MAE of the median of raw and post-processed irradiance forecasts and (b) difference in MAE of the median forecasts from the raw WRF ensemble as functions of the lead time.}
\label{fig:mae_maed}
\end{figure}

Finally, the MAE of the median of all five investigated predictions (Figure \ref{fig:mae_maed}a) and the difference in MAE of the median from the raw WRF ensemble (Figure \ref{fig:mae_maed}b), %as functions of the lead time, 
convey the same message as Figure \ref{fig:crps_crpss}. Compared to the raw ensemble, post-processing substantially improves the accuracy of the forecast median, and the corrected ensemble results in the lowest MAE values, followed by the DRN and EMOS approaches. Note that the medians of forecasts provided as full predictive distributions (CN0 EMOS and CN0 DRN) differ from the medians of their sample counterparts (CN0 EMOS-Q and CN0 DRN-Q); nevertheless, these differences are tiny as there are hardly any visible dissimilarities in the  MAE values of the matching predictions.

The above results indicate that all investigated post-processing approaches improve the calibration of probabilistic and the accuracy of point forecasts, and the corrected ensemble exhibits the best overall performance. The CN0 DRN model outperforms its EMOS counterpart in terms of both the mean CRPS and the MAE of the forecast median, but it is slightly more underdispersive, implying lower coverage and higher reliability indices.

\section{Conclusions}
\label{sec5}

We investigate the skill of probabilistic solar irradiance forecasts for the Atacama and Coquimbo regions in Chile. This area plays a crucial role in photovoltaic power production in the country, and our results can help obtain accurate probabilistic power forecasts. For the sake of this study, the Advanced Research module of the  Weather Research and Forecasting (WRF) model was utilized to generate short-term ensemble forecasts of solar irradiance for 30 locations with forecast horizons ranging up to 48 h with a temporal resolution of one hour. The forecasts comprise eight members differing in the planetary boundary layer and radiation parametrization. When verified against station observations, the raw ensemble forecasts exhibit a systematic positive bias for all forecast horizons and are underdispersive; however, the magnitude of the forecast error substantially varies for the different locations. 

For post-processing, we consider the parametric EMOS and DRN approaches based on a normal distribution left censored at zero and a nonparametric ensemble correction, where a neural network is trained to produce improved ensemble predictions. While all investigated post-processing methods substantially improve the calibration of probabilistic forecasts and reduce the MAE of the forecast median, this latter approach shows the best overall performance. From the competing parametric methods, the machine learning-based DRN outperforms the corresponding EMOS model, which aligns with the findings of \citet{bb24} and \citet{hkl24}.

Although this first study demonstrates the weaknesses of WRF irradiance ensemble forecasts and the potential in their statistical post-processing, there is still space for further improvements in calibration. On the one hand, one might try to augment the input features of the machine learning-based forecasts with predictions of related quantities. We have performed tests where WRF ensemble forecasts of temperature and wind speed were also included; however, the gain in forecast skill was minor, if any. On the other hand, one can test advanced neural network-based distribution-free methods such as the Bernstein quantile network \citep{brem20}, where the forecast distribution is modelled with a linear combination of Bernstein polynomials, or the histogram estimation network investigated, for instance, in \citet{sswh20}. 

Another interesting direction would be the investigation of various multivariate post-processing methods to produce temporary consistent forecast trajectories. Here, one can think of both the state-of-the-art two-step approaches such as the ensemble copula coupling \citep{stg13} or the Schaake shuffle \citep{cghrw04}, where the CN0 EMOS-Q, the CN0 DRN-Q or the corrected ensemble can serve as initial independent forecasts, and the recent data-driven conditional generative model of \citet{cjsl24}.

\bigskip
\noindent
{\bf Acknowledgments.} \ The authors gratefully acknowledge the support of the S\&T cooperation program 2021-1.2.4-T\'ET-2021-00020. S\'andor Baran, M\'aria Lakatos, and Marianna Szab\'o were also supported by the  National Research, Development, and Innovation Office under Grant No. K142849. Julio C\'esar Mar\'in and Omar Cuevas acknowledge the support of the Center of Atmospheric Studies and Climate Change of the University of Valpara\'\i so, Chile.

\end{document}